\documentclass[journal]{IEEEtran}
\IEEEoverridecommandlockouts

\usepackage{cite}
\usepackage{amsmath,amssymb,amsfonts,empheq,bm}
\usepackage{algorithmic}
\usepackage{graphicx}
\usepackage{textcomp}
\usepackage{xcolor}
\usepackage{float}

\usepackage{dotlessi}
\usepackage{layouts}
\usepackage{mathtools}
\usepackage{subcaption}
\usepackage[font=small]{caption}
\usepackage{multirow}

\newcommand{\ov}[1]{\overline{#1}}
\newcommand{\sss}{\scriptscriptstyle \rm}

\def\BibTeX{{\rm B\kern-.05em{\sc i\kern-.025em b}\kern-.08em
    T\kern-.1667em\lower.7ex\hbox{E}\kern-.125emX}}
    
\begin{document}

\title{Modeling Hybrid AC/DC Power Systems with the Complex Frequency
  Concept}

\author{Ignacio Ponce, \IEEEmembership{Member, IEEE}, and Federico
  Milano, \IEEEmembership{Fellow, IEEE}%
  \thanks{I.~Ponce and F.~Milano are with School of Electrical and
    Electronic Engineering, University College Dublin, Belfield
    Campus, D04V1W8, Ireland.  e-mails:
    \mbox{ignacio.poncearancibia@ucdconnect.ie},
    \mbox{federico.milano@ucd.ie}}%
  \thanks{This work is supported by the Sustainable Energy Authority
    of Ireland (SEAI) by funding I.~Ponce and F.~Milano under project
    FRESLIPS, Grant No. RDD/00681.}%
  \vspace{-0.6cm} }%

\maketitle

\begin{abstract}
  The concept of complex frequency has been recently introduced on the
  IEEE Transactions on Power Systems to study bus voltage variations
  in magnitude and frequency and their link with complex power
  injections of a power system.  In this paper, the complex frequency
  is applied to time-varying series connections, namely, RLC dynamic
  branches, regulating transformers and AC/DC converters.  The
  proposed modeling approach allows deriving an explicit expression
  for the complex frequency of the voltage of a certain bus as a
  linear combination of three elements: net current injected by the
  devices connected to the bus, adjacent voltages, and time-varying
  series branches.  The proposed formulation unifies the link between
  voltage and frequency dynamics in AC, DC, as well as hybrid AC/DC
  power systems.  A variety of static and dynamic examples are
  presented to show the potential of the proposed formulation.
  Relevant applications of the proposed modeling approach are
  outlined.
\end{abstract}


\section{Introduction}

\subsection{Motivation}

The dynamic behavior of power systems is experiencing unprecedented
changes due to the increasing penetration of converter-interfaced
devices.  Several recent works have discussed the challenges this
transition is posing to the modeling, control, and stability analysis
of power systems \cite{TaskForce, foundationsandchallenges, roleofCIG,
  SHAZON20226191, dorfler2023control}.  Among these challenges, the
most relevant for this paper is the need to revisit some fundamental
aspects of the modeling of power systems.  In particular, recent works
have highlighted the relevance of a more accurate definition and
interpretation of the frequency of power systems \cite{kirkham,frigo,
  milano2020frequency}.  A novel quantity, the complex frequency (CF),
has been introduced recently in \cite{ComplexFreq} precisely to
provide a more consistent foundation for the analysis of frequency
variations in AC power systems.  The most relevant feature of the CF
is the ability to link the variations of the complex power and the
variations of the voltage in magnitude and frequency in a common
framework.  This paper focuses on elaborating on the use of this
concept and further exploiting its potential.

\subsection{Literature Review}

The complex frequency has been used recently in promising applications
in power systems modeling.  For example, it is shown in
\cite{buttnercomplex} that the differential-algebraic equations that
describe the grid dynamics can be rewritten solely in terms of the CF
and state variables, removing completely the dependence on voltage
magnitudes and angles.  CF has also already found relevant
applications in control \cite{cfvfeedback, moutevelis}, converter
synchronization \cite{he2022complex, dorflerstability}, and state
estimation \cite{vppweilin, ZHONGinest}.  A relevant example is
provided in \cite{TaxonomyofPowerConverters}, where the concept is
used to describe and classify different control schemes and
synchronization mechanisms of power converters.  Although originally
applied to voltages and currents, the CF concept can be also applied
to any time-varying complex quantity.  In this paper, we define the CF
of time-varying admittances, which allows describing any series
element that can be modeled through an equivalent circuit.

The relationship between the CF at a bus and the complex powers was
originally presented in \cite{ComplexFreq} through an implicit
equation, where two types of contributions are distinguished.  On one
side of the equation, the terms correspond to voltage variations and
complex powers coming from neighbor buses connected through a constant
admittance, and they are said to be the network contribution.  On the
other side, the CF of the bus and the variations of the complex power
of every other element, referred to as the device's contribution.
Under this formulation, active series devices whose model is different
from a constant admittance block are treated similarly to shunt
devices instead of branches.  Thus, the effect of neighbor buses
connected through regulating transformers, AC/DC converters and other
special series connections cannot be directly evaluated.  The
extension of the use of the CF to admittances allows a more general
and consistent formulation where the effect of neighbor buses is
represented in the same way whether they are connected through an
active series device or not.  This approach allows us deriving an
explicit equation for the CF of the voltage of a bus with separate
terms for the effect of local shunt devices, the network, and dynamic
connections.

The aforementioned approach is particularly useful to study the
dynamic behavior of AC/DC converters.  In particular, we analyze the
dynamic link between AC- and DC-side dynamics, which depend on the
controllers configured on the converter.  This study is relevant as
the dynamic interaction of the converter's AC and DC circuits has been shown
to have a significant impact on the dynamic performance of the system.
For instance, authors in \cite{Mourouvin} show that for wind farms
connected through an MMC-HVDC interconnection, depending on the
control mode, there might be a threat to the stability of the DC side
and undesirable oscillations on the AC side.  The formulation proposed
in this work treats the converter as any other series branch, thus
giving the ability to relate the CF of the voltage at the AC and DC
sides using a single framework.

\subsection{Contributions}
The two main contributions of the paper are as follows:
\begin{itemize}
\item A systematic approach to model, based on the concept of CF, the
  dynamic behavior of different time-varying branches.
\item An explicit expression of the CF of the voltage of every bus of
  the system as a linear function of the net current injected by shunt
  devices at the bus, the effect of time-varying branches connected to
  the bus, and the CF of the voltage of neighbor buses.
\end{itemize}

The proposed framework is \textit{general}, as it unifies the modeling
of AC, DC or hybrid AC/DC power systems.  The approach is also
\textit{systematic} as long as the series element can be modeled
through an admittance block.  Finally, the formulation of the bus CFs
is \textit{exact} and does not require any assumption on the model of
the devices connected to the branch.

\subsection{Paper Organization}

The remainder of this paper is organized as follows.  Section
\ref{sec:background} provides a general definition and the notation
utilized in the reminder of the paper.  Section \ref{sec:formulation}
presents the proposed explicit equation for the complex
frequency. Specific expressions for the admittance model and the CF of
different types of time-varying branches are presented in Section
\ref{sec:timevarbranches}.  The implementation of the proposed
formulation in three study cases is described in Section
\ref{sec:studycases}, and some remarks on its potential applications
are provided in Section \ref{sub:remarks}.  Finally, Section
\ref{sec:conclusion} presents the conclusions and proposes future
work.

\section{Background}
\label{sec:background}

The complex frequency is, in turn, a derivative operator of a complex
number.  In fact, any complex quantity, say $\ov{u}$, with non-null
magnitude, i.e., $|\ov{u}| = u \ne 0$, can be written as:
\begin{equation}
   \ov{u} = {\rm exp}(\kappa + j \theta) \, ,
\end{equation}
where $\kappa = \ln (u)$.  If $\kappa$ and $\theta$ are smooth
functions of time, the time derivative of $\ov{u}$ gives:
\begin{equation}
  \dot{\ov{u}} = \frac{d \ov{u}}{dt} =
  (\dot{\kappa} + j \dot{\theta}) \, {\rm exp}(\kappa + j \theta) =
  \left (\frac{\dot{u}}{u} + j \dot{\theta} \right ) \, \ov{u} \, .
\end{equation}
In the reminder of this paper, the quantity
$(\frac{\dot{u}}{u} + j \dot{\theta})$ is called \textit{complex
  frequency}.  We define and utilize the CF of several quantities,
including voltages, currents and admittances.  The following notation
is used:
\begin{itemize}
\item $\ov{\eta}$ is the CF of voltage Park vectors,\footnote{The
    interested reader can find a discussion on the definition of Park
    vectors, which, in turn, are sort of dynamic phasors in
    \cite{milano2020frequency}.} namely
  $\dot{\ov{v}} = \ov{\eta} \, \ov{v}$.  For economy of notation, we
  also define $\ov{\eta} = \rho + j \omega$, where $\rho = \dot{v}/v$
  and $\omega = \dot{\theta}$.
\item $\ov{\xi}$ is the CF of current Park vectors, namely
  $\dot{\ov{\imath}} = \ov{\xi} \, \ov{\imath}$.
\item $\ov{\chi}$ is the CF of time-varying admittances, namely
  $\dot{\ov{Y}} = \ov{\chi} \, \ov{Y}$.
\end{itemize}
It can be shown that the CF is an invariant geometrical quantity, that
is, its value is the same independently from the reference frame
utilized to define the Park vector \cite{GeometricalFreq}.

It may be useful to note that the CF is a derived physical quantity.
It has thus nothing to do with the complex variable $s$ of the Laplace
transform that is widely utilized to study the frequency response of
circuits and control systems.  As a matter of fact, CF can be defined
both in the time-domain and in the $s$-domain of the Laplace
transform.  In the remainder of this work, we consider only the
time-domain.

\section{Proposed Formulation}
\label{sec:formulation}

In this section, we derive an expression for the CF of the voltage of
every bus as a linear combination of the CFs of neighbor voltages, the
current injection at the bus, and the admittance connected to the bus.

Consider a network with $n$ buses. We denote the set of the network
buses as $\mathbb{B}$ and the subset of buses containing every bus
except one bus $h$ as $\mathbb{B}_h=\mathbb{B} \setminus \{h\}$.
The starting point is the equation of the current balance of a node
$h \in \mathbb{B}$:
\begin{align}
  \ov{\dotlessi}_{h}
  & = \sum_{k\in \mathbb{B}_h}\ov{\dotlessi}_{h\rightarrow k}
    =\sum_{k\in \mathbb{B}_h}(\ov{v}_h-\ov{v}_k)\ov{Y}_{hk} \, ,
    \label{eq:current_balance1}
\end{align}
where $\ov{\dotlessi}_h$ is the net current injection at bus $h$ from
every device connected to $h$.  It is important to remark that there
is no assumption on the expression of $\ov{Y}_{hk}$.  It can represent
the admittance of any series element, dynamic or not, connected
between nodes $h$ and $k$.

Isolating $\ov{v}_h$ from
(\ref{eq:current_balance1}):
\begin{equation}
  \label{eq:current_balance2}
  \ov{v}_h\sum_{k\in \mathbb{B}_h}\ov{Y}_{hk} =
  \sum_{k\in \mathbb{B}_h}\ov{v}_k\ov{Y}_{hk}+\ov{\dotlessi}_h \, .
\end{equation}

We denote $\ov{Y}_{hh}=-\sum_{k\in \mathbb{B}_h}\ov{Y}_{hk}$ similarly
to how the diagonal elements of the admittance matrix of a power
system are defined.  Thus (\ref{eq:current_balance2}) can be
equivalently written as:
\begin{equation}
  \label{eq:current_balance3}
  -\ov{v}_h\ov{Y}_{hh} =
  \sum_{k\in \mathbb{B}_h}\ov{v}_k\ov{Y}_{hk}+\ov{\dotlessi}_h \, .
\end{equation}

Differentiating (\ref{eq:current_balance3}) and recalling the property
of the CF to act as a linear derivative operator \cite{ComplexFreq}:
\begin{equation}
  \label{eq:current_balance2dt}
  -\ov{v}_h\ov{Y}_{hh}(\ov{\eta}_h+\ov{\chi}_{hh}) =
  \sum_{k\in \mathbb{B}_h}\ov{v}_k\ov{Y}_{hk}(\ov{\eta}_k+\ov{\chi}_{hk})
  +\ov{\dotlessi}_h\ov{\xi}_h \, ,
\end{equation}
where $\ov{\eta}_h$ is the CF of the voltage of bus $h$; $\ov{\xi}_h$
is the CF of the net current injected at the same bus;
$\ov{\chi}_{hk}$ is the CF of the admittance $\ov{Y}_{hk}$; and
$\ov{\chi}_{hh}$ is the CF of the admittance $\ov{Y}_{hh}$, equal to:
\begin{equation}
    \ov{\chi}_{hh} = -\frac{\sum_{k\in \mathbb{B}_h}\ov{Y}_{hk}\ov{\chi}_{hk}}{\ov{Y}_{hh}} \, .
\end{equation}

Solving (\ref{eq:current_balance2dt}) for $\ov{\eta}_h$ and
rearranging some terms:
\begin{equation}\label{eq:cfd_base_ew0}
  \ov{\eta}_h=\sum_{k\in \mathbb{B}_h}\frac{(\ov{v}_h-\ov{v}_k)\ov{Y}_{hk}\ov{\chi}_{hk}}{\ov{v}_h\ov{Y}_{hh}}-\sum_{k\in \mathbb{B}_h}\frac{\ov{v}_k\ov{Y}_{hk}\ov{\eta}_k}{\ov{v}_h\ov{Y}_{hh}}-\frac{\ov{\dotlessi}_h\ov{\xi}_h}{\ov{v}_h\ov{Y}_{hh}} \, .
\end{equation}

Recalling (\ref{eq:current_balance1}), (\ref{eq:cfd_base_ew0})
becomes:
\begin{equation}
  \label{eq:cfd_base_ew1}
  \ov{\eta}_h = \sum_{k\in \mathbb{B}_h} \frac{\ov{\dotlessi}_{h\rightarrow k}
    \ov{\chi}_{hk}}{\ov{v}_h\ov{Y}_{hh}} - \sum_{k\in \mathbb{B}_h}
  \frac{\ov{v}_k\ov{Y}_{hk}\ov{\eta}_k}{\ov{v}_h\ov{Y}_{hh}}
  - \frac{\ov{\dotlessi}_h\ov{\xi}_h}{\ov{v}_h\ov{Y}_{hh}} \, .
\end{equation}

Let us define the quantities $\ov{c}_{\chi_{hk}}$, $\ov{c}_{\eta_{hk}}$ and $\ov{c}_{\xi_h}$ as:
\begin{align}
    \ov{c}_{\chi_{hk}} = \frac{\ov{\dotlessi}_{h\rightarrow k}}{\ov{v}_h\ov{Y}_{hh}} \, , \; \;
    \ov{c}_{\eta_{hk}} = -\frac{\ov{v}_k\ov{Y}_{hk}}{\ov{v}_h\ov{Y}_{hh}} \, , \; \; 
  \ov{c}_{\xi_h} = \frac{-\ov{\dotlessi}_h}{\ov{v}_h\ov{Y}_{hh}} \, .
  \label{eq:cfd_coeffs_def_xi}
\end{align}
Then, (\ref{eq:cfd_base_ew1}) leads to the sought expression of the CF
of the bus voltages:
\begin{equation}
  \label{eq:cfd_base_ew2}
  \boxed{\ov{\eta}_h = \sum_{k\in \mathbb{B}_h}\ov{c}_{\chi_{hk}}\ov{\chi}_{hk} +
    \sum_{k\in \mathbb{B}_h}\ov{c}_{\eta_{hk}}\ov{\eta}_k + \ov{c}_{\xi_h}\ov{\xi}_h}
\end{equation}

In (\ref{eq:cfd_base_ew2}), the CF of the voltage of a certain bus $h$
is expressed as a linear combination of the CF of three different
types of variables: admittance connected to $h$, weighted by a complex
coefficient $\ov{c}_{\chi_{hk}}$, adjacent voltages, weighted by a
complex coefficient $\ov{c}_{\eta_{hk}}$, and the net current injected
at the node, weighted by the complex coefficient $\ov{c}_{\xi_h}$.

%
The coefficients of (\ref{eq:cfd_base_ew2}) can be interpreted as a
measure of the participation of three elements to the dynamic of the
voltage: series branches ($\ov{c}_{\chi_{hk}}$), neighboring bus
voltages ($\ov{c}_{\eta_{hk}}$) and net current injections (
$\ov{c}_{\xi_h}$).  Valuable information can be extracted by analyzing
the characteristics of these complex coefficients.  For example, the
real part of $\ov{c}_{\eta_{hk}}$ can be viewed as a measure of the
self-participation of neighbor voltages on the CF of every bus.  In
other words, the real part of $\ov{c}_{\eta_{hk}}$ measures how much
$\rho_k$ affects $\rho_h$ and $\omega_k$ affects $\omega_h$.  On the
other hand, the imaginary part of $\ov{c}_{\eta_{hk}}$ represents the
cross-participation, i.e., how much neighbor $\rho_k$ affects
$\omega_h$ and $\omega_k$ affects $\rho_h$.  Hence, it is to be expect
that the real part of $\ov{c}_{\eta_{hk}}$ is greater than its
imaginary part, i.e., a small $\rho\leftrightarrow\omega$ dynamic
coupling compared to the $\rho\leftrightarrow\rho$ and
$\omega\leftrightarrow\omega$ direct link between two consecutive
buses.
Moreover, we note that all the coefficients in (\ref{eq:cfd_base_ew2})
are dimensionless since they relate normalized variations of different
variables expressed in terms of CFs.  While the concept of the CF of
the current $\ov{\xi}$ and the voltage $\ov{\eta}$ has been
comprehensively explained in \cite{ComplexFreq}, the value of
$\ov{\chi}$ depends on what is actually being modeled as a
time-varying admittance.  For instance, dynamic series devices such as
regulating transformers, AC/DC converters, or simply RLC
dynamics. These examples are studied in the following section.
To complete this section, we note that, for constant admittances,
(\ref{eq:cfd_base_ew2}) reduces to:
\begin{equation}\label{eq:cfd_base_noserdyn}
  \ov{\eta}_h = \sum_{k\in \mathbb{B}_h}\ov{c}_{\eta_{hk}}
  \ov{\eta}_k + \ov{c}_{\xi_h}\ov{\xi}_h \, .
\end{equation}
Then, for buses without a shunt-connected device, namely, transit
buses, (\ref{eq:cfd_base_ew2}) reduces to a linear combination of only
neighbor buses dynamics:
\begin{equation}
  \label{eq:cfd_base_transit}
  \ov{\eta}_h=\sum_{k\in \mathbb{B}_h}\ov{c}_{\eta_{hk}}\ov{\eta}_k \, .
\end{equation}

\section{Modeling of Time-varying branches}
\label{sec:timevarbranches}

Specific expressions for $\ov{\chi}$ are derived in this section
for different cases of dynamic admittances.

\subsection{Dynamic RLC Branches}

\begin{figure}[H]
  \centering
  \includegraphics{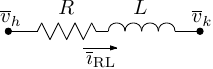} \hspace{0.65cm}
  \includegraphics{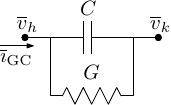}
  \caption{Left: series RL circuit; right: parallel GC circuit.}
  \label{fig:RL}
\end{figure}

Consider the circuit shown in the left side of Fig.~\ref{fig:RL}.

The well-known equation that links the current and the voltage applied
to a series RL circuit is:
\begin{equation}
  L \, \dot{\ov{\dotlessi}}_{\sss RL} +
  R \, \ov{\dotlessi}_{\sss RL} = \ov{v}_h-\ov{v}_k \, .
\end{equation}

Using the CF as a derivative operator:
\begin{align}
    L \, \ov{\dotlessi}_{\sss RL}\ov{\xi}+R \, \ov{\dotlessi}_{\sss RL}&=\ov{v}_h-\ov{v}_k \, , \\
    \ov{\dotlessi}_{\sss RL}\left(L \, \ov{\xi}+R\right)&=\ov{v}_h-\ov{v}_k \, , \\
    \ov{\dotlessi}_{\sss RL}&=\frac{1}{L \, \ov{\xi}+R}\left(\ov{v}_h-\ov{v}_k\right) \, , \\
    \Rightarrow \ov{Y}_{\sss RL}&=\frac{1}{L \, \ov{\xi}+R} \, .\label{eq:yrl}
\end{align}

Thus, $\ov{\chi}_{\sss RL}$ can be written as:
\begin{align}
  \ov{\chi}_{\sss RL} = \frac{\dot{\ov{Y}}_{\sss RL}}{\ov{Y}_{\sss RL}}\quad
  \Rightarrow \quad \Aboxed{\ov{\chi}_{\sss RL} =
  -\frac{\dot{\ov{\xi}}}{\ov{\xi}+\frac{R}{L}}}\label{eq:chi_RL}
\end{align}

Following a similar procedure, it can be shown that the admittance of
a dynamic parallel GC branch, as shown in the right circuit of
Fig.~\ref{fig:RL}, is:
\begin{equation}
  \ov{Y}_{\sss GC}=C\ov{\eta}+G \, ,
\end{equation}
where $\ov{\eta}$ is the CF of the voltage applied between both
terminals of the GC block, i.e., $\left(\ov{v}_h-\ov{v}_k\right)$.
Finally, $\ov{\chi}_{\sss GC}$ is obtained as:
\begin{align}
  \ov{\chi}_{\sss GC} = \frac{\dot{\ov{Y}}_{\sss GC}}{\ov{Y}_{\sss GC}} \quad
  \Rightarrow \quad \Aboxed{\ov{\chi}_{\sss GC} =
  \frac{\dot{\ov{\eta}}}{\ov{\eta}+\frac{G}{C}}}\label{eq:chi_GC}
\end{align}

For lossless circuits, (\ref{eq:chi_RL}) and (\ref{eq:chi_GC}) become:
\begin{align}
  \ov{\chi}_{\sss RL}|_{R=0} =-\frac{\dot{\ov{\xi}}}{\ov{\xi}} \, ,
  \quad \text{and} \quad
  \ov{\chi}_{\sss GC}|_{G=0} =\frac{\dot{\ov{\eta}}}{\ov{\eta}} \, .
\end{align}

The CF of RLC series connections, PI-type lines, or any other dynamic
circuit composed of a combination of these basic blocks can be derived
using the equations provided in this subsection.


\subsection{Regulating Transformer}
\label{subsec:ret}

\begin{figure}[H]
    \centering
    \includegraphics{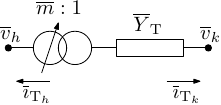}
    \caption{Regulating transformer equivalent circuit.}
    \label{fig:PhST}
\end{figure}

Regulating transformers are series-connected devices that can modify
the magnitude and-or the phase angle difference between two nodes of a
meshed network.  Regardless its control scheme, the regulating
transformer can be modeled as a series of an ideal transformer with a
variable complex tap ratio and an admittance \cite{kundur}.
A graphical representation of the circuit is shown in
Fig.~\ref{fig:PhST}.

In Fig.~\ref{fig:PhST}, $\ov{m}=me^{j\alpha}$ represents the complex
tap ratio and $\ov{Y}_{\sss T}$ is the admittance of the transformer.
The link between the terminal voltages and currents can be written in
terms of a non-symmetric admittance matrix $\ov{\mathbf Y}_{\sss T}$
as follows \cite{kundur}:
\begin{align}
  \begin{bmatrix}
    \ov{\dotlessi}_{\mathrm{T}_k} \\
    \ov{\dotlessi}_{\mathrm{T}_h}
  \end{bmatrix}
  =
  \begin{bmatrix}
    -\ov{Y}_{\sss T} & me^{j\alpha}\ov{Y}_{\sss T} \\
    me^{-j\alpha}\ov{Y}_{\sss T} & -m^2\ov{Y}_{\sss T}
  \end{bmatrix}
  \begin{bmatrix}
    \ov{v}_{k} \\
    \ov{v}_{h}
  \end{bmatrix} =
  \ov{\mathbf Y}_{\sss T}
  \begin{bmatrix}
    \ov{v}_{k} \\
    \ov{v}_{h}
  \end{bmatrix} .
\end{align}

Then, $\ov{\chi}$ is calculated for each term of the admittance
matrix of the regulating transformer by imposing the following
equation:
\begin{equation}
  \label{eq:chi_PhST}
  \dot{\ov{\mathbf Y}}_{\sss T}={\ov{\mathbf X}}_{\sss T} 
  \circ {\ov{\mathbf Y}}_{\sss T} \, ,
\end{equation}
where the dot over the matrix denotes the time derivative of each
matrix element, and $\circ$ represents the Hadamard product, i.e., the
element-by-element product of the two matrices; and
$\ov{\mathbf X}_{\sss T}$ is:
\begingroup
\renewcommand*{\arraystretch}{2.5}
\begin{equation}
\boxed{
\ov{\mathbf X}_{\sss T}=
\begin{bmatrix}
    0 & \dfrac{\dot{m}}{m}+j\dot{\alpha}\\
    \dfrac{\dot{m}}{m}-j\dot{\alpha} & 2 \, \dfrac{\dot{m}}{m}
\end{bmatrix}}
\end{equation}
\endgroup

\subsection{AC-DC Converter}
\label{subsec:acdcconv}

\begin{figure}[htb]
    \centering
    \includegraphics{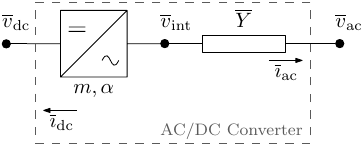}
    \caption{AC/DC converter graphical representation.}
    \label{fig:ac_dc_converter}
\end{figure}

Figure \ref{fig:ac_dc_converter} shows a typical scheme of an AC/DC
converter, which includes a bidirectional ideal converter and a
transformer on the AC side of the converter.

Note that $\ov{v}_{\rm dc}$ and $\ov{\imath}_{\rm dc}$ are assumed to
be complex quantities, although their phase angle is null, to allow
for a general formulation of hybrid AC/DC systems. Neglecting the
magnetization and the iron losses of the transformer, the electrical
equations of the averaged model of the converter are
\cite{MilanoOrtegaStorage}:
\begin{align}
    \ov{v}_{\rm int}&=\ov{v}_{\rm dc}\ov{m} \, , \label{eq:acdc} \\
    \ov{m}&=me^{j(\theta_{\rm ac}+\alpha)} \, , \label{eq:mpolar}\\
    \ov{\dotlessi}_{\rm ac}&=\left(\ov{v}_{\rm int}-\ov{v}_{\rm ac}\right)\ov{Y} \, , \label{eq:acdc3}\\
    0&=\Re\{\ov{v}_{\rm dc}\ov{\dotlessi}_{\rm dc}\}+\Re\{\ov{v}_{\rm int}\ov{\dotlessi}_{\rm ac}^{*}\} \, ,\label{eq:acdc4}
\end{align}
where, using the notation of Fig.~\ref{fig:ac_dc_converter}, $m$ is
the scaling factor of the AC/DC converter; $\alpha$ is the phase shift
the AC/DC converter introduces between AC node
$\ov{v}_{\rm ac} = v_{\rm ac} \angle \theta_{\rm ac}$ and the internal
node $\ov{v}_{\rm int}$; and $\ov{Y}$ is the series admittance of the
filter and transformer on the AC side of the converter. Note that,
while the rectangular form is the most common representation of
$\ov{m}$ in the literature, i.e., $\ov{m}=m_d+jm_q$, we have
nevertheless utilized the polar form shown in (\ref{eq:mpolar}) for
consistency with the formulation of the other models.

It is important to remark that the derivation that follows is valid
independently from the control scheme implemented in the converter.
This is due to the fact that the admittance block and, thus,
$\ov{\chi}$, depend only on the electrical equations of the AC/DC
converter (\ref{eq:acdc})-(\ref{eq:acdc4}).  Thus, the dynamics of the
controllers are implicitly reflected in $\ov{m}$.

Differentiating $\ov{\chi}$ requires prior knowledge of the equivalent
admittance block of the converter. One approach to obtain this block
is to first represent the AC/DC converter model as an equivalent
circuit, which requires including additional elements to be consistent
with (\ref{eq:acdc})-(\ref{eq:acdc4}). Examples of this strategy can
be found in \cite{Rezvani} and \cite{Acha}, where a shunt susceptance
is included to balance the reactive power.  An alternative, more
direct approach, which we use in this work, is to manipulate
(\ref{eq:acdc})-(\ref{eq:acdc4}) to obtain the following form:
\begin{equation}\label{eq:iyv_desired}
\begin{bmatrix}
    \ov{\dotlessi}_{\rm ac} \\
    \ov{\dotlessi}_{\rm dc}
\end{bmatrix}=
    \begin{bmatrix}
         \ov{Y}_{\rm acac}& \ov{Y}_{\rm acdc}\\{}
         \ov{Y}_{\rm dcac}& \ov{Y}_{\rm dcdc}
    \end{bmatrix}
\begin{bmatrix}
    \ov{v}_{\rm ac} \\
    \ov{v}_{\rm dc}
\end{bmatrix} .
\end{equation}
By taking the first row of (\ref{eq:iyv_desired}):
\begin{equation}\label{eq:acdc1row}
  \ov{\dotlessi}_{\rm ac}=\ov{Y}_{\rm acac}\ov{v}_{\rm ac}+\ov{Y}_{\rm acdc}\ov{v}_{\rm dc} \,.
\end{equation}
On the other hand, from (\ref{eq:acdc3}) and using (\ref{eq:acdc}):
\begin{equation}\label{eq:acdc1rowmodel}
  \ov{\dotlessi}_{\rm ac}=-\ov{Y}\ov{v}_{\rm ac}+\ov{m}\ov{Y}\ov{v}_{\rm dc}\,.
\end{equation}
Comparing (\ref{eq:acdc1row}) and (\ref{eq:acdc1rowmodel}), $\ov{Y}_{\rm acac}$ and $\ov{Y}_{\rm acdc}$ are found:
\begin{align}
    \ov{Y}_{\rm acac}&=-\ov{Y} \, , \\
    \ov{Y}_{\rm acdc}&=me^{j\left(\alpha+\theta_{\rm ac}\right)}\ov{Y}\, .  
\end{align}
Next, by taking the second row of (\ref{eq:iyv_desired}) and splitting the real and imaginary part:
\begin{align}
  \dotlessi_{\rm dc}&=\Re\{\ov{Y}_{\rm dcac}\}\Re\{\ov{v}_{\rm ac}\}-\Im\{\ov{Y}_{\rm dcac}\}\Im\{\ov{v}_{\rm ac}\}+\Re\{\ov{Y}_{\rm dcdc}\}v_{\rm dc}\, ,\label{eq:acdc2rowre}\\
  0&=\Re\{\ov{Y}_{\rm dcac}\}\Im\{\ov{v}_{\rm ac}\}+\Im\{\ov{Y}_{\rm dcac}\}\Re\{\ov{v}_{\rm ac}\}+\Im\{\ov{Y}_{\rm dcdc}\}v_{\rm dc}\, . \label{eq:acdc2rowim}
\end{align}
On the other hand, from (\ref{eq:acdc4}) and using (\ref{eq:acdc}) and (\ref{eq:acdc3}):
\begin{align}
  0&=v_{\rm dc}\dotlessi_{\rm dc}+v_{\rm dc}\Re\{\ov{m}\ov{\dotlessi}_{\rm ac}^{*}\} \, ,\\
  \dotlessi_{\rm dc}&=-\Re\left\{\ov{m}(\ov{v}_{\rm int}^{*}-\ov{v}_{\rm ac}^{*})\ov{Y}^{*}\right\} \, ,\\
  \dotlessi_{\rm dc}&=-\Re\left\{\ov{m}^{*}(\ov{m}v_{\rm dc}-\ov{v}_{\rm ac})\ov{Y}\right\} \, ,\\
  \dotlessi_{\rm dc}&=\Re\left\{-m^{2}\ov{Y}\right\}v_{\rm dc}+\Re\left\{\ov{Y}\ov{m}^{*}\ov{v}_{\rm ac}\right\}\, ,\\
  \dotlessi_{\rm dc}&=\Re\left\{\ov{Y}\ov{m}^{*}\right\}\Re\{\ov{v}_{\rm ac}\}-\Im\left\{\ov{Y}\ov{m}^{*}\right\}\Im\{\ov{v}_{\rm ac}\}+\dots\label{eq:acdc2rowmodelre}\\
  &\dots+\Re\left\{-m^{2}\ov{Y}\right\}v_{\rm dc}\, .\nonumber 
\end{align}
Considering $\ov{Y}=Ye^{j\gamma}=G+jB$, and comparing (\ref{eq:acdc2rowre}) and (\ref{eq:acdc2rowmodelre}):
\begin{align}\label{eq:acdc2rowreres}
  \ov{Y}_{\rm dcac}&= \ov{Y}\ov{m}^{*}=me^{-j\left(\alpha+\theta_{\rm ac}\right)}\ov{Y}\, ,\\
  \Re\{\ov{Y}_{\rm dcdc}\}&=\Re\left\{-m^{2}\ov{Y}\right\} = -m^2G \, .
\end{align}
Replacing (\ref{eq:acdc2rowreres}) in (\ref{eq:acdc2rowim}):
\begin{align}
  0&=\Im\{\ov{Y}\ov{m}^{*}\ov{v}_{\rm ac}\}+\Im\{\ov{Y}_{\rm dcdc}\}v_{\rm dc}\,, \\
  \Im\{\ov{Y}_{\rm dcdc}\}&=-\Im\left\{\frac{\ov{Y}\ov{m}^{*}\ov{v}_{\rm ac}}{v_{\rm dc}}\right\}\,,\\
  \Im\{\ov{Y}_{\rm dcdc}\}&=jm\frac{v_{\rm ac}}{v_{\rm dc}}Y\sin(\alpha-\gamma)\,,\\
  \Im\{\ov{Y}_{\rm dcdc}\}&=jm\frac{v_{\rm ac}}{v_{\rm dc}}(G\sin(\alpha)-B\cos(\alpha))\,.
\end{align}
Finally, the equivalent admittance block of the AC/DC converter is
obtained as follows:
\begin{equation}
    \mathbf{\ov{Y}_{\rm \mathbf{\rm acdc}}}=
    \begin{bmatrix}
         \ov{Y}_{\rm acac}& \ov{Y}_{\rm acdc}\\
         \ov{Y}_{\rm dcac}& \ov{Y}_{\rm dcdc}
    \end{bmatrix} ,
\end{equation}
where:
\begin{align}
    \ov{Y}_{\rm acac}&=-\ov{Y} \, , \\
    \ov{Y}_{\rm acdc}&=me^{j\left(\alpha+\theta_{\rm ac}\right)}\ov{Y} \, , \\
    \ov{Y}_{\rm dcac}&=me^{-j\left(\alpha+\theta_{\rm ac}\right)}\ov{Y} \, , \\
    \ov{Y}_{\rm dcdc}&=-m^2G+jm\frac{v_{\rm ac}}{v_{\rm dc}}\left(G\sin(\alpha)-B\cos(\alpha)\right) \, .
\end{align}
Then, $\ov{\chi}$ is calculated for each term of
$\ov{\mathbf Y}_{\rm acdc}$ by imposing:
\begin{equation}\label{eq:chi_acdc}
  \dot{\ov{\mathbf Y}}_{\rm \mathbf{\rm acdc}} =
  \ov{\mathbf X}_{\rm acdc}\circ \ov{\mathbf Y}_{\rm acdc} \, .
\end{equation}

The calculation of $\ov{\chi}_{\rm acac}$, $\ov{\chi}_{\rm acdc}$,
$\ov{\chi}_{\rm dcac}$ is analogous to the calculations of the
coefficients of the regulating transformer:
\begin{align}
    \Aboxed{\ov{\chi}_{\rm acac}&=0}\\
    \Aboxed{\ov{\chi}_{\rm acdc}&=\frac{\dot{m}}{m}+j\left(\dot{\alpha}+\dot{\theta}_{\rm ac}\right)}\\
    \Aboxed{\ov{\chi}_{\rm dcac}&=\frac{\dot{m}}{m}-j\left(\dot{\alpha}+\dot{\theta}_{\rm ac}\right)}
\end{align}

On the other hand, the term $\ov{\chi}_{\rm dcdc}$ requires more work. First, $\ov{Y}_{\rm dcdc}$ is separated as the following:
\begin{align}
    \ov{Y}_{\rm dcdc}&=\ov{Y}_{\rm dcdc}^{(1)}+\ov{Y}_{\rm dcdc}^{(2)} \, ,\\
    \ov{Y}_{\rm dcdc}^{(1)}&=-m^2G \, , \\
    \ov{Y}_{\rm dcdc}^{(2)}&=jm\frac{v_{\rm ac}}{v_{\rm dc}}\left(G\sin(\alpha)-B\cos(\alpha)\right) \, , \\
    \Rightarrow \quad \ov{\chi}_{\rm dcdc}&=\frac{\ov{Y}_{\rm dcdc}^{(1)}\ov{\chi}_{\rm dcdc}^{(1)}+\ov{Y}_{\rm dcdc}^{(2)}\ov{\chi}_{\rm dcdc}^{(2)}}{\ov{Y}_{\rm dcdc}^{(1)}+\ov{Y}_{\rm dcdc}^{(2)}} \, ,
\end{align}
where
\begin{align}
   \ov{\chi}_{\rm dcdc}^{(1)}&= 2 \, \frac{\dot{m}}{m} \, , \\
   \ov{\chi}_{\rm dcdc}^{(2)}&=\frac{\dot{m}}{m}+\frac{\dot{v}_{\rm ac}}{v_{\rm ac}}-\frac{\dot{v}_{\rm dc}}{v_{\rm dc}}+\dot{\alpha}\frac{G\cos(\alpha)+B\sin(\alpha)}{G\sin(\alpha)-B\cos(\alpha)} \, .
\end{align}

If losses are neglected, $\ov{\chi}_{\rm dcdc}$ becomes:
\begin{equation}
  \boxed{\ov{\chi}_{\rm dcdc}|_{G=0} =
    \dfrac{\dot{m}}{m} + \dfrac{\dot{v}_{\rm ac}}{v_{\rm ac}} -
    \dfrac{\dot{v}_{\rm dc}}{v_{\rm dc}} - \dot{\alpha}\tan(\alpha)}
\end{equation}

\section{Case Studies}
\label{sec:studycases}

In this section, we apply the proposed general expression
\eqref{eq:cfd_base_ew2} for the CF as well as the various
model-dependent expressions derived in Section
\ref{sec:timevarbranches} to three benchmark systems.  In Section
\ref{sub:ac}, we use a purely AC system to present a steady-state
analysis of the values of the coefficients of (\ref{eq:cfd_base_ew2}).
In Section \ref{sub:dc}, we use a DC network to present a dynamic
analysis focused on the coefficients of our equation regarding dynamic
RLC branches.  In Section \ref{sub:acdc}, we present a hybrid system
including different types of time-varying branches described in
Section \ref{sec:timevarbranches}, namely RLC dynamic branches and
AC/DC converters.  We also show the ability of our general formulation
to study together such a diverse system and the different ways the
AC/DC converter propagates the frequency dynamics depending on its
control mode.  Finally, Section \ref{sub:remarks} collects remarks on
practical applications of the proposed modeling approach.

\subsection{AC System}
\label{sub:ac}

\begin{figure}[htb]
  \centering
  \includegraphics[scale=0.9]{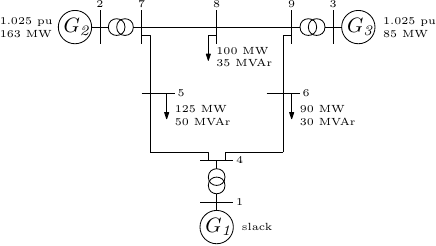}
  \caption{Single-line diagram of the WSCC 9-bus system.}
  \label{fig:wscc}
\end{figure}

The proposed formulation is applied to the WSCC 9-bus benchmark
system.  The network consists of three synchronous generators feeding
three loads through a delta-type topology.  The single-line diagram of
the WSCC system is shown in Fig.~\ref{fig:wscc}.

\begin{table*}[t]
  \caption{Steady state coefficients of the CF for the WSCC 9-bus system.}
  \begin{center}
{
\renewcommand{\arraystretch}{1.5}
\begin{tabular}{c|cccccccccc}
\hline
{\textbf{Bus}} & $\bm{\ov{c}_{\eta_{h\mathit{1}}}}$           & $\bm{\ov{c}_{\eta_{h\mathit{2}}}}$          & $\bm{\ov{c}_{\eta_{h\mathit{3}}}}$           & $\bm{\ov{c}_{\eta_{h\mathit{4}}}}$           & $\bm{\ov{c}_{\eta_{h\mathit{5}}}}$        & $\bm{\ov{c}_{\eta_{h\mathit{6}}}}$           & $\bm{\ov{c}_{\eta_{h\mathit{7}}}}$           & $\bm{\ov{c}_{\eta_{h\mathit{8}}}}$           & $\bm{\ov{c}_{\eta_{h\mathit{9}}}}$          & $\bm{\ov{c}_{\xi_{h}}}$ \\ \hline
\textbf{1}  &   -          &    -         &      -       & 0.99-j0.04   &   -       &       -      &    -         &    -         &    -         & 0.01+j0.04               \\ 
\textbf{2}  &    -         &     -        &     -        &      -       &  -        &       -      & 1.00-j0.10       & -            & -            & 0.00+j0.10                     \\ 
\textbf{3}  & -            & -            & -            & -            & -         & -            & -            & -            & 1.01-j0.05 & -0.01+j0.05              \\ 
\textbf{4}  & 0.45-j0.02 &-           &-           &-           & 0.29+j0.00 & 0.27+j0.02 &-           &-           &-           &-                       \\ 
\textbf{5}  &-           &-           &-           & 0.69+j0.00    &-        &-           & 0.35+j0.07 &-           &-           & -0.04-j0.07              \\ 
\textbf{6}  &-           &-           &-           & 0.67+j0.01 &-        &-           &-           &-           & 0.36+j0.04 & -0.03-j0.05              \\ 
\textbf{7}  &-           & 0.45+j0.01 &-           &-           & 0.17+j0.00 &-           &-           & 0.38-j0.01 &-           &-                       \\ 
\textbf{8}  &-           &-           &-           &-           &-        &-           & 0.59+j0.03 &-           & 0.43+j0.01   & -0.02-j0.04              \\
\textbf{9}  &-           &-           & 0.53-j0.02 &-           &-        & 0.17+j0.01 &-           & 0.30+j0.01  &-           &-                       \\ \hline
\end{tabular}}
\label{tab:wscc_c0}
\end{center}
\end{table*}

Let us consider first the case of an admittance matrix with constant
elements.  Thus, we use the simplified expression
(\ref{eq:cfd_base_noserdyn}).  The power flow solution is found for
the base-case operating conditions shown in Fig.~\ref{fig:wscc}.  The
solution is used to calculate $\ov{c}_{\eta_{hk}}$ and
$\ov{c}_{\xi_{h}}$ for every bus $h$.  The steady-state coefficients
given by the power flow solution are shown in Table \ref{tab:wscc_c0}.

Columns 2-10 of the table show the participation of neighbor buses to
the CF of every bus.  Elements corresponding to non-neighbor buses are
null, as expected, and left blank in the table.  The table is sparse
in the same way the admittance matrix of a system is sparse, i.e.,
rather than an absolute measure of the relationship between every pair
of buses, the information has the same structure of the connectivity
matrix of the grid.  Thus, even if a coefficient linking a given pair
of buses is null, the voltage variations in one bus impact on the
voltage of the other bus though the intermediate bus connections.  The
coefficients $\ov{c}_{\eta_{hk}}$ are predominantly real, which means
that, as expected, the results verify a small
$\rho \leftrightarrow \omega$ dynamic coupling compared to the
$\rho \leftrightarrow \rho$ and $\omega \leftrightarrow \omega$ direct
link.  The impact of the net injected current is shown in the last
column of Table \ref{tab:wscc_c0}.  As this coefficient is
proportional to the net current injected at the bus, buses with a
shunt device interchanging more power exhibit a higher magnitude than
those with less power.  For instance, the magnitude of
$\ov{c}_{\xi_{h}}$ for bus 2, which has $G_{\it 2}$ injecting 163 MW,
is approximately twice compared to the value for bus 3, which has
$G_{\it 3}$ injecting 85 MW.  In the case of transit buses 4, 7 and 9,
$\ov{c}_{\xi_{h}}$ is null.  However, the coefficients
$\ov{c}_{\xi_{h}}$ are generally much smaller than the coefficients
$\ov{c}_{\eta_{h}}$ because of the different scaling between voltage
and current coefficients.  Thus, the different magnitudes are not an
indication that the current injected by devices has a negligible
impact on the voltage.

\textit{Property 1:} It is interesting to note that a remarkable
property of the coefficients is verified: for each bus, each row in
Table \ref{tab:wscc_c0} always sums exactly $1$.  The proof can be
derived by taking \eqref{eq:current_balance2} and dividing both sides
by the left-hand side of the equation.  This property has a physical
interpretation: for every bus, the sum of the self-participation of
neighbor buses and the net current injected naturally adds up to
$100 \%$, whereas the sum of the cross-participations vanishes.

Finally, as a sensibility analysis, the line connecting buses 7 and 8
is replaced by a regulating transformer to examine its corresponding
coefficient according to (\ref{eq:cfd_base_ew2}).  The model used is
as described in \ref{subsec:ret}, with a series resistance and
inductance equal to the parameters of the original line.  The tap
ratio is set to $1$, and the active power reference is set equal to
the solution of the power flow of the original case.  Under these
conditions, the coefficients presented in Table \ref{tab:wscc_c0} are
unaltered.  However, the coefficients modeling the dynamic effect of
the regulating transformer also become relevant for buses 7 and 8.
The resulting expression for the CF of bus 7 is:

\begin{align}
  \ov{\eta}_7 = \underbrace{\sum_{k\in \{2, 5, 8\}} \ov{c}_{\eta_{7k}} \ov{\eta}_k}_{\substack{\text{Original terms}}} +
  \underbrace{\ov{c}_{\chi_{78}} \left(\frac{\dot{m}}{m} + j\dot{\alpha}\right)}_{\substack{\text{Regulating transformer term}}}
\end{align}
where $\ov{c}_{\chi_{78}} = -(0.01+0.03j)$.  Note that
$\ov{c}_{\chi_{78}}$, which weights the impact of the control of the
regulating transformer on $\ov{\eta}_7$ has a small magnitude, and has
same order of magnitude of the coefficients $\ov{c}_{\xi}$ of the net
injected current.  However, since the control of the regulating
transformer is ``slow'', one has to expect that its effect on
$\ov{\eta}_7$ is small.


\subsection{DC System}
\label{sub:dc}

\begin{figure}[htb]
  \centering
  \includegraphics{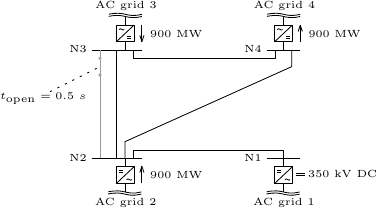}
  \caption{Scheme of the DC subsystem of the Chaudhuri's MTDC system.}
  \label{fig:mtdc}
\end{figure}

The proposed expression \eqref{eq:cfd_base_ew2} is utilized for the DC
subsystem of the hybrid Chaudhuri's Multi-terminal DC (MTDC) system
\cite{chaudhuri}.  This subsystem consists of a 4-bus 350 kV DC grid
with four converter stations, each connected to an equivalent AC grid.
The scheme of the DC subsystem is shown in Fig.~\ref{fig:mtdc}.  The
original data are modified by including an auxiliary RL branch between
nodes N2 and N3 with ten times the resistance and inductance of the
existing connection.  The perturbation consists in disconnecting the
auxiliary RL branch.

The DC transmission system is modeled considering series resistance,
inductance, and shunt capacitance of all branches.  A time-domain
simulation is carried out to study how the coefficients of
(\ref{eq:cfd_base_ew2}), especially those capturing line dynamic
effects, vary during the transient.

\begin{figure}[h]
  \centering
  \includegraphics[scale=0.95]{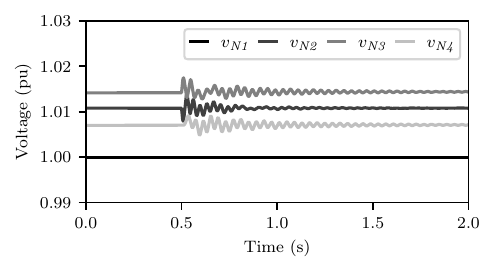}
  \caption{Trajectories of the voltages of DC nodes.}
  \label{fig:mtdc_vh}
\end{figure}

Starting from the operating point given by the conditions shown in
Fig.~\ref{fig:mtdc}, the auxiliary RL branch is disconnected at
$t=0.5$ s.  The trajectories of the voltages of the four DC nodes is
shown in Fig.~\ref{fig:mtdc_vh}.  The perturbation causes an
oscillation of about 27 Hz which is due to the coupling of branch
inductances and capacitances.  These oscillations are noticeable on
the DC voltage of every bus except N1.  This happens because the
$d$-axis control loop configured in converter one actively controls
the DC voltage with a faster time constant than the dynamics triggered
by the perturbation.

\begin{figure}[h]
  \centering
  \includegraphics[scale=0.95]{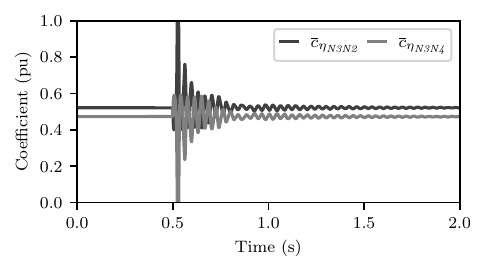}
  \caption{Trajectories of the real part of the coefficients
    corresponding to the neighbor voltages for N3.}
  \label{fig:mtdc_cb_03_0}
\end{figure}

\begin{figure}[h]
  \centering
  \includegraphics[scale=0.95]{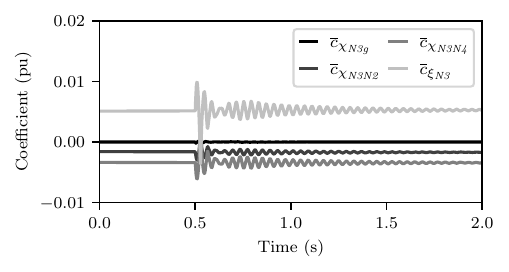}
  \caption{Trajectories of the real part of the coefficients
    corresponding to dynamic branches and net current injected
    coefficients for N3.}
  \label{fig:mtdc_cb_03_1}
\end{figure}

The trajectories of the real part of the coefficients of
(\ref{eq:cfd_base_ew2}) for N3 is shown in
Figs.~\ref{fig:mtdc_cb_03_0} and \ref{fig:mtdc_cb_03_1}.  The
imaginary part is null for the DC grid.  As expected, since all the
coefficients depend on the system variables, they inherit the dynamic
triggered by the perturbation.
Figure \ref{fig:mtdc_cb_03_1} indicates that the coefficient related
to the dynamic of shunt capacitance blocks
($\ov{c}_{\chi_{\mathit{N3g}}}$) is considerably lower and almost
negligible than those corresponding to series inductance blocks.  This
difference is due to the current flowing through each branch.

\textit{Property 2:} It is also interesting to observe another
property of the proposed expression \eqref{eq:cfd_base_ew2}: the sum
of all the coefficients corresponding to the time-varying branches
($\ov{c}_{\chi_{\mathit{N3N2}}}+\ov{c}_{\chi_{\mathit{N3N4}}}$) equals
the value of the net injected current coefficient
($\ov{c}_{\xi_{\mathit{N3}}}$).  This property can also be obtained
from \eqref{eq:current_balance1} and the definition of the
coefficients $\ov{c}_{\chi_{hk}}$ given in
\eqref{eq:cfd_coeffs_def_xi}.

Finally, note that $\ov{c}_{\eta_{\mathit{N3N2}}}$,
$\ov{c}_{\eta_{\mathit{N3N4}}}$ and $\ov{c}_{\xi_{\mathit{N3}}}$
exhibit a spike in the second oscillation peak after the perturbation.
The rationale behind this phenomenon lies in the fact that the
modeling approach presented in Section \ref{sec:timevarbranches} uses
instantaneous equivalent admittance blocks that depend on variable CF
quantities.  In particular, (\ref{eq:yrl}) shows that for an RL
branch, the equivalent admittance equals the inverse of
$(L\ov{\xi}+R)$.  Consequently, it is possible to have a singularity
in case $(L\ov{\xi}+R)=0$. In the case of AC grids, the issue does not
occur due to the imaginary part of $\ov{\xi}$, which corresponds to
the instantaneous frequency of the current.  However, in DC grids,
$(L\ov{\xi}+R)$ is real and can be viewed as an equivalent resistance
with expression:
\begin{equation}
  \label{eq:req_sing}
  R_{\rm eq}= L \, \frac{\dot{\dotlessi}}{\dotlessi}+R \, .
\end{equation}

\begin{figure}[h]
  \centering
  \includegraphics[scale=0.95]{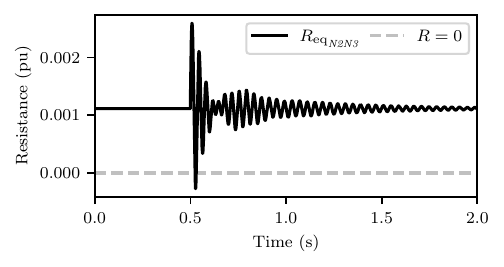}
  \caption{Trajectories of the equivalent resistance of branch N2-N3.}
  \label{fig:mtdc_z}
\end{figure}

To illustrate this peculiar behavior, Fig.~\ref{fig:mtdc_z} shows the
equivalent resistance $R_{\rm eq}$ of branch N2-N3, which change sign
in correspondence of the spike of the trajectories shown in
Fig.~\ref{fig:mtdc_cb_03_0}.

\subsection{Hybrid AC/DC System}
\label{sub:acdc}

\begin{figure}[htb]
  \centering
  \includegraphics[scale=0.9]{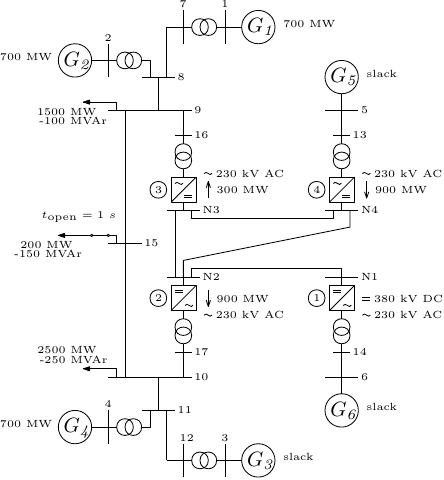}
  \caption{Single-line diagram of the Chaudhuri's MTDC system.}
  \label{fig:chaudhuri}
\end{figure}

Here we illustrate the behavior of the elements of the proposed
formula (\ref{eq:cfd_base_ew2}) for the complete hybrid Chaudhuri's
MTDC system \cite{chaudhuri}.
A multi-terminal DC (MTDC) system is connected to the grid through two
different buses.  The DC grid is also connected to two remote isolated
AC grids.  The main AC grid consists of a multi-machine AC network
with two areas, two synchronous generators each.  The MTDC system
injects real power into the main AC grid. The system's single-line
diagram is shown in Fig.~\ref{fig:chaudhuri}.

The DC lines are modeled considering series resistance, inductance,
and shunt capacitance, as described in Section \ref{sub:dc}.  For
simplicity but without lack of generality, the AC transmission system
is modeled as a constant admittance matrix.  Besides, the four AC/DC
converters are modeled as in Section \ref{subsec:acdcconv} plus
different control schemes. Their controlled variables are shown in
Table \ref{tab:cmodes} for each device. While converters 2 and 3
contribute to the frequency regulation in the main AC area, converter
4 supports the frequency of the isolated AC grid it is connected to.
Converter 1 sustains the DC voltage instead of having a
frequency-control loop.  Finally, all the synchronous generators are
equipped with standard governors and AVR models.  There is also an AGC
in the three AC areas.

\begin{table}[htb]
  \centering
  \caption{Control modes configured in each converter.}
  \renewcommand{\arraystretch}{1.25}
  \begin{tabular}{c|cccc}
    \hline
    \multirow{2}{*}{Control Loop} & \multicolumn{4}{c}{Converter} \\
                                  & 1 & 2 & 3 & 4 \\
    \hline
    $d$-axis (P-control)  & $v_{\rm dc}$ & $f_{\rm ac}$ & $f_{\rm ac}$ & $f_{\rm ac}$ \\ 
    $q$-axis (Q-control)  & $v_{\rm ac}$ & $v_{\rm ac}$ & $v_{\rm ac}$ & $v_{\rm ac}$ \\
    \hline
  \end{tabular}
  \label{tab:cmodes}
\end{table}

\begin{figure}[h]
  \centering
  \includegraphics[scale=0.95]{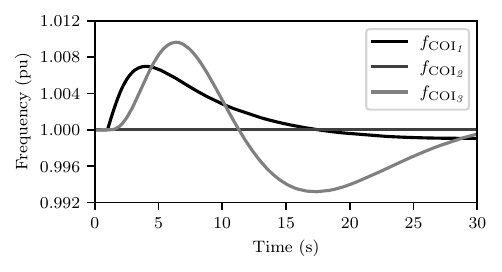}
  \caption{Trajectories of the COI-frequencies of the three AC areas
    of the Chaudhuri's MTDC system.}
  \label{fig:ch_omega}
\end{figure}

\begin{figure}[h]
  \centering
  \includegraphics[scale=0.95]{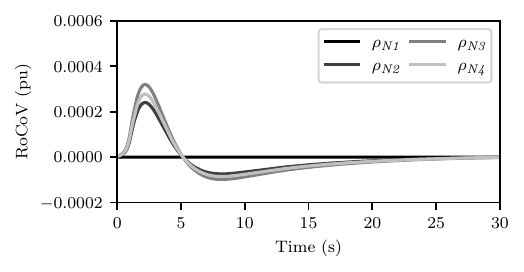}
  \caption{Trajectories of the DC RoCoVs of the Chaudhuri's MTDC
    system.}
  \label{fig:ch_vdc}
\end{figure}

A time-domain simulation is carried out to study the relationship
between the CF of AC and DC buses based on the proposed equation.  In
particular, we show the effect of the AC/DC converter model depending
on the control mode configured on each converter. The contingency is a sudden disconnection of the load connected at bus
15 at $t=1$ s.  Figures \ref{fig:ch_omega} and \ref{fig:ch_vdc} show
the evolution of the frequency of the COI of the three AC areas and
the DC RoCoV at the four DC nodes, respectively.  As expected, the
frequency of the main AC area ($f_{\rm COI_{\it 1}}$) increases after
the load disconnection.  Consequently, converters 2 and 3 reduce their
power injection, leading to a deviation in the DC voltage at N2 and
N3.  The AC dynamic propagates through the DC network to converters 1
and 4, which show different behaviors.  Converter 4 observes a dynamic
event at N4, but it does not propagate it to the AC side, i.e.,
$f_{\rm COI_{\it 2}}$ is constant.  On the other hand, converter 1
keeps a constant DC voltage at N1 ($\rho_{\mathit{N1}}=0$), but
current variations propagate the dynamic event to the AC side, leading
to a significant frequency deviation in that area
($f_{\rm COI_{\it 3}}$).  The difference lies in the dynamics of $m$
and $\alpha$ according to the control mode of each converter.

Equation \eqref{eq:cfd_base_ew2} is implemented at buses 13 and 14 to
explain the effect of the control of converters 1 and 4. Since our
formulation allows us to treat the AC/DC converters as branches, buses
13 and 14 become transit buses, and thus, \eqref{eq:cfd_base_ew2}
becomes:

\begin{equation}\label{eq:cfd_ch1314}
  \ov{\eta}_h = \sum_{k\in \mathbb{B}_h}\ov{c}_{\chi_{hk}}\ov{\chi}_{hk} +
  \sum_{k\in \mathbb{B}_h}\ov{c}_{\eta_{hk}}\ov{\eta}_k
\end{equation}

\begin{figure}[h]
  \centering
  \includegraphics[scale=0.8]{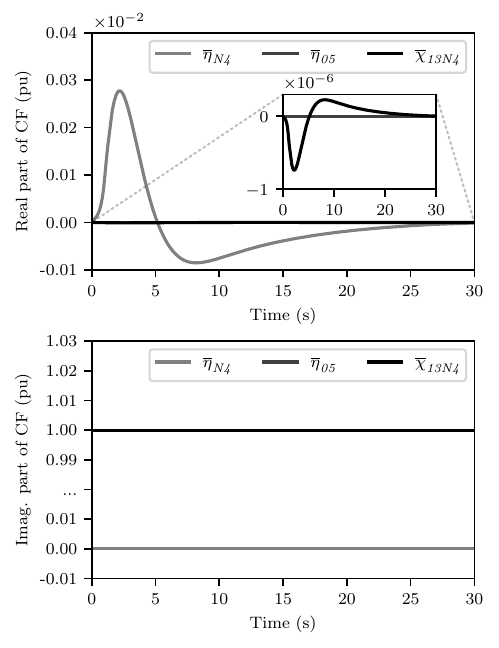}
  \caption{Trajectories of the CFs of non-null terms of
    (\ref{eq:cfd_ch1314}) for bus 13 of the Chaudhuri's MTDC system.}
  \label{fig:ch_analysis_13}
\end{figure}

In the case of bus 13, the only non-null terms of
\eqref{eq:cfd_ch1314} are those corresponding to adjacent buses:
$\ov{\eta}_{\it 05}$, $\ov{\eta}_{\it N4}$, and the AC-DC converter:
$\ov{\chi}_{\it 13N4}$.  The time evolution of the real and imaginary
parts of these three CFs is shown in Fig.~\ref{fig:ch_analysis_13}.
The dynamic caused by the contingency is seen by converter 4 at the DC
bus in the real part of $\ov{\eta}_{\it N4}$.  The imaginary part is
zero by definition (DC bus).  The reaction of the converter is
expressed through $\ov{\chi}_{\it 13N4}$.  Despite the different
magnitudes, the real part of $\ov{\chi}_{\it 13N4}$ exhibits the same
dynamic but with an opposed sign.

\begin{figure}[h]
  \centering
  \includegraphics[scale=0.8]{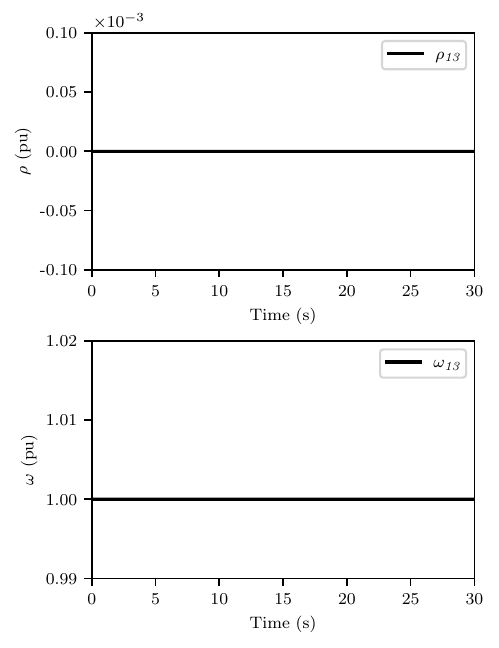}
  \caption{Trajectories of the CF at bus 13 of the Chaudhuri's MTDC
    system. Calculated using equation (\ref{eq:cfd_base_ew2}).}
  \label{fig:ch_eta26_13}
\end{figure}

The adjacent AC bus remains unaltered, i.e., it shows a zero real part
and a constant unit imaginary part.  The contribution of the three CFs
is multiplied by the corresponding complex coefficients and added
according to (\ref{eq:cfd_ch1314}) to give $\ov{\eta}_{\it 13}$.  The
time evolution of $\ov{\eta}_{\it 13}$ is shown in
Fig.~\ref{fig:ch_eta26_13}.  The effect of the converter is such that
it compensates the dynamic received at $N4$, thus giving an unaltered
$\ov{\eta}$ at bus 13, i.e., zero real part and unit imaginary part.
In other words, the DC dynamic does not propagate to the AC side.

\begin{figure}[h]
  \centering
  \includegraphics[scale=0.8]{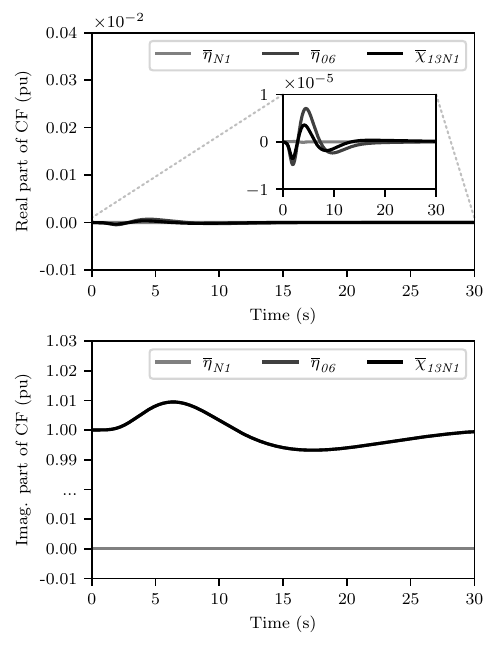}
  \caption{Trajectories of the CFs of non-null terms
    of (\ref{eq:cfd_ch1314}) for bus 14 of the Chaudhuri's MTDC
    system.}
  \label{fig:ch_analysis_14}
\end{figure}

In the case of bus 14, the only non-null terms of
(\ref{eq:cfd_ch1314}) are those corresponding to adjacent buses:
$\ov{\eta}_{\it 06}$, $\ov{\eta}_{\it N1}$, and the AC-DC converter:
$\ov{\chi}_{\it 14N1}$.  The trajectories of the real and imaginary
parts of these three CFs are shown in Fig. \ref{fig:ch_analysis_14}.
In this case, the real part of the CF at the DC bus
$\ov{\eta}_{\it N1}$ is nearly zero as converter 1 controls the DC
voltage.  However, this requires the converter to move its control
variables, thus inducing a dynamic in both the real and imaginary
parts of $\ov{\chi}_{\it 13N1}$, which is finally propagated to the AC
side. Fig. \ref{fig:ch_eta26_13} shows the result for
$\ov{\eta}_{\it 14}$.  In this case, the converter does propagate the
dynamic to the AC side.

\begin{figure}[h]
  \centering
  \includegraphics[scale=0.8]{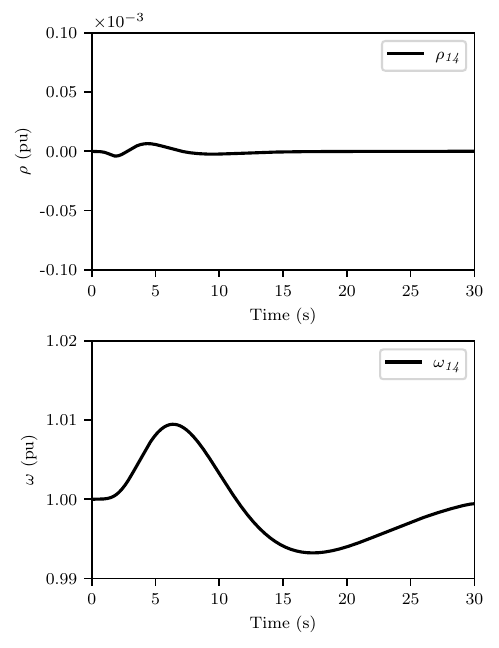}
  \caption{Trajectories of the CF at bus 14 of the Chaudhuri's MTDC
    system. Calculated using equation (\ref{eq:cfd_base_ew2}).}
  \label{fig:ch_eta26_14}
\end{figure}

\subsection{Remarks and Applications}
\label{sub:remarks}
The steady-state analysis of the coefficients of the proposed
expression \eqref{eq:cfd_base_ew2} given in Section \ref{sub:ac}
illustrates how these coefficients are useful dimensionless weights of
the participation of adjacent buses to the dynamic of the voltage of a
particular bus.  Moreover, the ratio between the real and imaginary
parts indicates the instantaneous coupling between $\rho$ and
$\omega$.  On the one hand, estimating the value of the coefficients
appears useful for monitoring the power system state, especially for
weak interconnections.  On the other hand, taking their value as an
input for a control scheme arises as an application worth exploring.
Dynamic analysis shows that despite being affected by the dynamics of
the system, the steady-state value of the coefficients of the
expression tends to prevail.  This confirms the potential use of the
coefficients that appear in \eqref{eq:cfd_base_ew2} as a metric to
evaluate the level of synchronization between adjacent buses.

The formulation of series elements of the network as dynamic
admittances requires the definition of special quantities, which are
useful for measuring some phenomena present in the system.  For
instance, the instantaneous equivalent resistance carries the
information of the instantaneous energy exchange between $L$ and $R$
in an $RL$ branch.  The CFs of the dynamic admittances also appear as
useful quantities to evaluate the dynamic impact of time-varying
(e.g., controlled) series devices. 

The proposed equation offers a quantitative framework for the analysis
of time-domain simulation results. Particularly interesting is to
explore its ability to evaluate the interaction between dynamic
devices and the propagation of dynamics events. Notably, due to the
exact nature of the formulation, this capability is model-agnostic,
e.g. simulation of EMT or RMS models.

The dynamic admittance model used to represent AC/DC converters allows
unifying the formulation of hybrid systems.  Besides, the proposed
equation explicitly relates the dynamics of the AC and DC side
together with the effect of the converter.  This opens the possibility
of studying how to design a control scheme that achieves a specific
objective of the propagation of $\rho$ and $\omega$.  For instance,
designing a control that translates the dynamic of $\omega$ of the AC
side to $\rho$ of the DC side and blocks the dynamic of $\rho$ of the
AC side at the same time.  Such a scheme naturally transfer the
information of the state of the frequency of a remote AC area
throughout the DC grid without the need for communications.

\section{Conclusion}
\label{sec:conclusion}

The paper introduces an explicit equation for the CF of the voltage in
terms of the CF of three variables: the net current injected at the
bus, the voltage of adjacent buses, and dynamic branches connected to
the bus.  The paper also provides specific expressions for dynamic RLC
circuits, regulating transformers, and AC/DC converters.  The
representation of the AC/DC converter allows modeling hybrid systems
in a unique formulation, and keeps track of the relationship between
the dynamic behavior of the real and imaginary parts of the complex
frequencies of the AC side and the DC side.

The case studies show that the proposed formulation is a useful tool
to monitor the power system state, as the coefficients of
(\ref{eq:cfd_base_ew2}) constitute a dimensionless measure of the
dynamic link between adjacent buses.  The novel quantities required to
build the unique formulation presented in this paper can also be
applied to quantify transient phenomena, such as the instantaneous
energy exchange in a dynamic RL circuit or the reaction of active
series branches.  The latter case is especially promising in the case
of AC/DC converters, where the formulation opens the possibility of
designing a control scheme to achieve a specific objective of
$\rho\leftrightarrow\omega$ propagation from AC to DC and vice-versa.

Future work will focus on the extension of the expression
(\ref{eq:cfd_base_ew2}) to relevant devices, e.g., detailed models of
grid-following and grid-following converters.  Finally, upcoming
research will explore applications of (\ref{eq:cfd_base_ew2}) for the
monitoring, control and stability analysis of power
systems.

\begin{IEEEbiography}[{\includegraphics[width=1in, height=1.25in,
    clip, keepaspectratio]{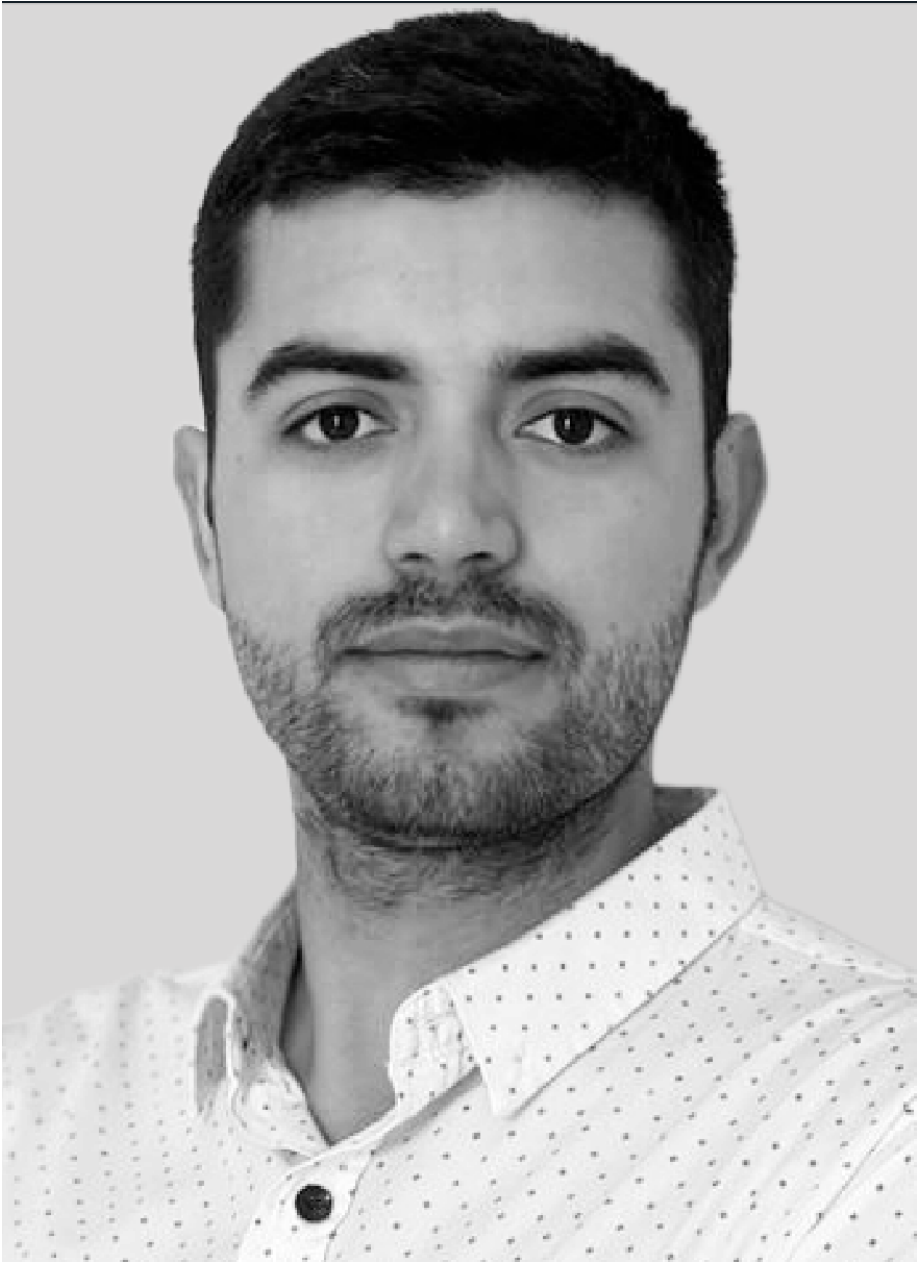}}] {Ignacio Ponce}
  received from University of Chile the BSc.~and MSc.~degree in
  Electrical Engineering in 2019 and 2022, respectively.  He is
  currently pursuing a Ph.D in Electrical Engineering at University
  College Dublin, Ireland.  His research interests include power
  system modeling, control and stability analysis.
\end{IEEEbiography}

\begin{IEEEbiography}[{\includegraphics[width=1in, height=1.25in,
    clip, keepaspectratio]{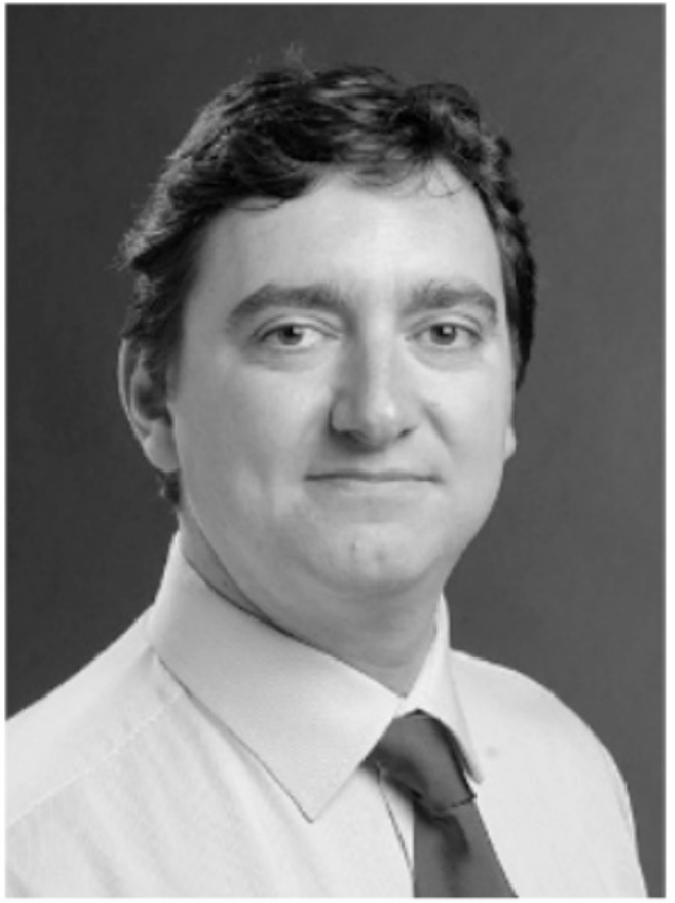}}] {Federico Milano}
  (F'16) received from the Univ.~of Genoa, Italy, the Ph.D.~in
  Electrical Engineering 2003.  In 2013, he joined the University
  College Dublin, Ireland, where he is currently a full professor.  He
  is Chair of the IEEE Power System Stability Controls Subcommittee,
  IET Fellow, IEEE PES Distinguished Lecturer, Chair of the Technical
  Program Committee of the PSCC 2024, Senior Editor of the IEEE
  Transactions on Power Systems, Member of the Cigr{\'e} Irish
  National Committee, and Co-Editor in Chief of the IET Generation,
  Transmission \& Distribution.  His research interests include power
  system modeling, control and stability analysis.
\end{IEEEbiography}

\vfill


\begin{thebibliography}{10}

\bibitem{TaskForce}
N.~Hatziargyriou {\em et~al.}, ``Definition and classification of power system
  stability – revisited \& extended,'' {\em IEEE Transactions on Power
  Systems}, vol.~36, no.~4, pp.~3271--3281, 2021.

\bibitem{foundationsandchallenges}
F.~Milano, F.~Dörfler, G.~Hug, D.~J. Hill, and G.~Verbič, ``Foundations and
  challenges of low-inertia systems (invited paper),'' in {\em Power Systems
  Computation Conference (PSCC)}, pp.~1--25, 2018.

\bibitem{roleofCIG}
J.~Fang, Y.~Tang, H.~Li, and F.~Blaabjerg, ``The role of power electronics in
  future low inertia power systems,'' in {\em IEEE International Power
  Electronics and Application Conference and Exposition (PEAC)}, pp.~1--6,
  2018.

\bibitem{SHAZON20226191}
M.~N.~H. Shazon, Nahid-Al-Masood, and A.~Jawad, ``Frequency control challenges
  and potential countermeasures in future low-inertia power systems: A
  review,'' {\em Energy Reports}, vol.~8, pp.~6191--6219, 2022.

\bibitem{dorfler2023control}
F.~D{\"o}rfler and D.~Gro{\ss}, ``Control of low-inertia power systems,'' {\em
  Annual Review of Control, Robotics, and Autonomous Systems}, vol.~6,
  pp.~415--445, 2023.

\bibitem{kirkham}
H.~Kirkham, W.~Dickerson, and A.~Phadke, ``Defining power system frequency,''
  in {\em IEEE PES General Meeting (PESGM)}, pp.~1--5, 2018.

\bibitem{frigo}
G.~Frigo, A.~Derviškadić, and M.~Paolone, ``Impact of fundamental frequency
  definition in ipdft-based pmu estimates in fault conditions,'' in {\em IEEE
  10th International Workshop on Applied Measurements for Power Systems
  (AMPS)}, pp.~1--6, 2019.

\bibitem{milano2020frequency}
F.~Milano and {\'A}.~Ortega, {\em Frequency Variations in Power Systems:
  Modeling, State Estimation, and Control}.
\newblock IEEE Press, Wiley, 2020.

\bibitem{ComplexFreq}
F.~Milano, ``Complex frequency,'' {\em IEEE Transactions on Power Systems},
  vol.~37, no.~2, pp.~1230--1240, 2022.

\bibitem{buttnercomplex}
A.~Büttner and F.~Hellmann, ``Complex couplings -- a universal, adaptive and
  bilinear formulation of power grid dynamics,'' 2023.

\bibitem{cfvfeedback}
F.~Milano, B.~Alhanjari, and G.~Tzounas, ``Enhancing frequency control through
  rate of change of voltage feedback,'' {\em IEEE Transactions on Power
  Systems}, pp.~1--4, 2023.

\bibitem{moutevelis}
D.~Moutevelis, J.~Rold{\'a}n-P{\'e}rez, M.~Prodanovic, and F.~Milano, ``Design
  of virtual impedance control loop using the complex frequency approach,'' in
  {\em 2023 IEEE Belgrade PowerTech}, pp.~1--6, IEEE, 2023.

\bibitem{he2022complex}
X.~He, V.~H{\"a}berle, and F.~D{\"o}rfler, ``Complex-frequency synchronization
  of converter-based power systems,'' {\em arXiv preprint arXiv:2208.13860},
  2022.

\bibitem{dorflerstability}
X.~He, V.~Häberle, I.~Subotić, and F.~Dörfler, ``Nonlinear stability of
  complex droop control in converter-based power systems,'' {\em IEEE Control
  Systems Letters}, vol.~7, pp.~1327--1332, 2023.

\bibitem{vppweilin}
W.~Zhong, G.~Tzounas, and F.~Milano, ``Real-time estimation of vpp equivalent
  inertia and fast frequency control,'' in {\em IEEE PES General Meeting
  (PESGM)}, pp.~1--5, 2022.

\bibitem{ZHONGinest}
W.~Zhong, G.~Tzounas, M.~Liu, and F.~Milano, ``On-line inertia estimation of
  virtual power plants,'' {\em Electric Power Systems Research}, vol.~212,
  p.~108336, 2022.

\bibitem{TaxonomyofPowerConverters}
D.~Moutevelis, J.~Roldán-Pérez, M.~Prodanovic, and F.~Milano, ``Taxonomy of
  power converter control schemes based on the complex frequency concept,''
  {\em IEEE Transactions on Power Systems}, pp.~1--13, 2023.

\bibitem{Mourouvin}
R.~Mourouvin, K.~Shinoda, J.~Dai, A.~Benchaib, S.~Bacha, and D.~Georges,
  ``{AC/DC} dynamic interactions of {MMC-HVDC} in grid-forming for wind-farm
  integration in {AC} systems,'' in {\em 22nd European Conference on Power
  Electronics and Applications (EPE'20 ECCE Europe)}, pp.~1--9, 2020.

\bibitem{GeometricalFreq}
F.~Milano, ``A geometrical interpretation of frequency,'' {\em IEEE
  Transactions on Power Systems}, vol.~37, no.~1, pp.~816--819, 2022.

\bibitem{kundur}
P.~Kundur, N.~Balu, and M.~Lauby, {\em Power System Stability and Control}.
\newblock EPRI power system engineering series, McGraw-Hill Education, 1994.

\bibitem{MilanoOrtegaStorage}
F.~Milano and {\'A}.~Ortega, {\em Converter-Interfaced Energy Storage Systems:
  Context, Modelling and Dynamic Analysis}.
\newblock Cambridge University Press, 2019.

\bibitem{Rezvani}
M.~M. Rezvani and S.~Mehraeen, ``A generalized model for unified ac-dc load
  flow analysis,'' in {\em 2021 IEEE Texas Power and Energy Conference (TPEC)},
  pp.~1--6, 2021.

\bibitem{Acha}
E.~Acha, B.~Kazemtabrizi, and L.~M. Castro, ``A new vsc-hvdc model for power
  flows using the newton-raphson method,'' {\em IEEE Transactions on Power
  Systems}, vol.~28, no.~3, pp.~2602--2612, 2013.

\bibitem{chaudhuri}
N.~Chaudhuri, B.~Chaudhuri, R.~Majumder, and A.~Yazdani, {\em Multi-terminal
  Direct-Current Grids: Modeling, Analysis, and Control}.
\newblock IEEE Press, Wiley, 2014.

\end{thebibliography}
\end{document}